%% file: sigchi.tex
\documentclass[sigchi, review=false]{acmart}

\renewcommand\footnotetextcopyrightpermission[1]{} 

\usepackage{booktabs} 

\usepackage{CJKutf8}
\newcommand{\zh}[1]{\begin{CJK}{UTF8}{gbsn}#1\end{CJK}}

\usepackage{array}
\newcolumntype{L}[1]{>{\raggedright\let\newline\\\arraybackslash\hspace{0pt}}m{#1}}
\newcolumntype{C}[1]{>{\centering\let\newline\\\arraybackslash\hspace{0pt}}m{#1}}
\newcolumntype{R}[1]{>{\raggedleft\let\newline\\\arraybackslash\hspace{0pt}}m{#1}}

\usepackage{color, colortbl}
\definecolor{Gray}{gray}{0.9}
\definecolor{White}{rgb}{1,1,1}

\usepackage[linesnumbered,ruled]{algorithm2e}

\SetCommentSty{mycommfont}

\acmConference{}
\acmYear{}
\copyrightyear{}

\begin{document}
\title[Engaging Audiences in Virtual Museums by Interactively Prompting Guiding Questions]{Engaging Audiences in Virtual Museums by Interactively Prompting Guiding Questions}

\author{Zhenjie Zhao}
\affiliation{%
  \institution{Hong Kong University of Science and Technology}
  \city{Hong Kong}
}
\email{zzhaoao@cse.ust.hk}

\begin{abstract}
    Virtual museums aim to promote access to cultural artifacts. However, they often face the challenge of getting audiences to read and understand a large amount of information in an uncontrolled online environment. Inspired by successful practices in physical museums, we investigated the possible use of guiding questions to engage audiences in virtual museums. To this end, we first identified how to construct questions that are likely to attract audiences through domain expert interviews and mining cultural-related posts in a popular question and answer community. Then in terms of the proactive level for attracting users' attention, we designed two mechanisms to interactively prompt questions: active and passive. Through an online experiment with 150 participants, we showed that having interactive guiding questions encourages browsing and improves content comprehension. We discuss reasons why they are useful by conducting another qualitative comparison and obtained insights about the influence of question category and interaction mechanism.
\end{abstract}

%
%




\maketitle

\input{body}




\end{document}

%% file: body.tex
\section{Introduction}

A virtual museum, or digital museum, collects, organizes, and displays 
digital artifacts online, which is mainly created for educational and entertainment 
purposes \cite{falk1998effect,vimmdefinition,Schweibenz1998TheM,carrozzino2018comparing}. 
Compared to a physical museum, a virtual museum is not limited by geographical locations,
and can provide richer user experience to fulfill its instructional and recreational function
\cite{SUN20131242,TheCountenanceofPublicCulturalFacilities}. 
A recent research reports that digital museum visitors have outnumbered those of 
real museums \cite{hawkey:hal-00190496}. 
This indicates the increasing popularity of the new form of 
interacting with cultural artifacts. 
In China for example, economic growth generates a new surge of public interest
in tangible and intangible cultural properties \cite{romero2014scholar}.
Chinese museums are thus motivated to cultivate curiosity about the history and 
diversity of China with the help of digital and web technologies \cite{romero2014scholar}.
For instance, the Palace Museum alone has digitalized about
one million artifacts into high-resolution images and
put them online for people to browse at will \cite{TheCountenanceofPublicCulturalFacilities}.

\begin{figure}[h]
  \centering
  \includegraphics[width=1.0\linewidth]{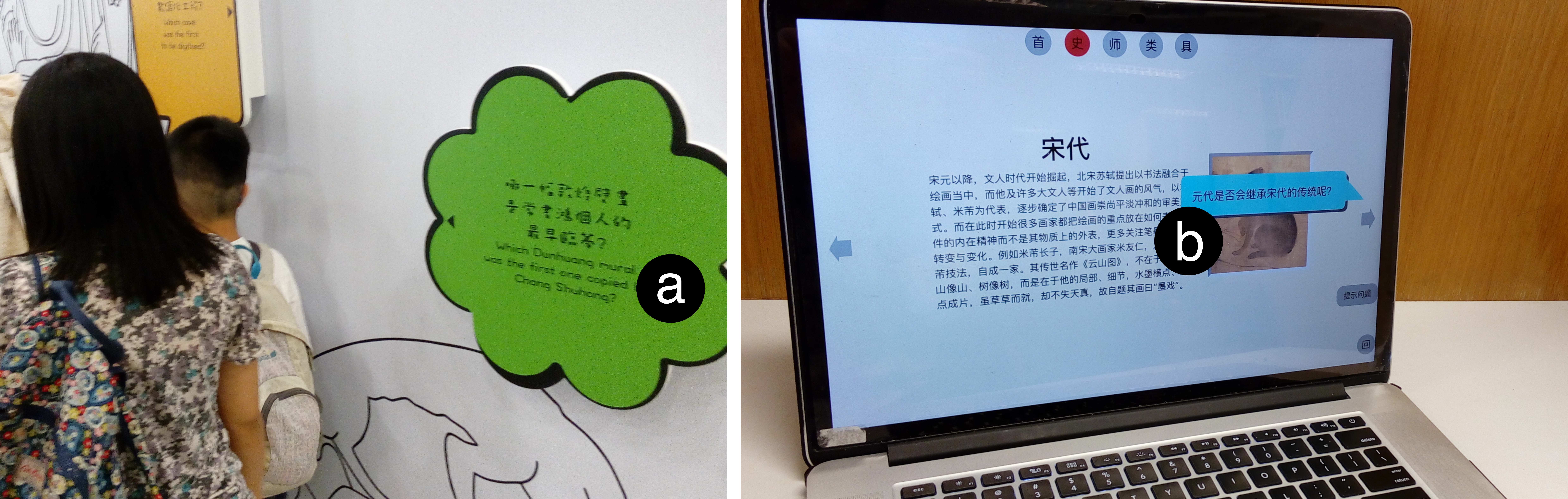}
  \caption{Left: guiding questions in an exhibition on Dunhuang Caves.
  Right: one exhibition in our study that uses interactive guiding questions.
  ((a). ``Which Dunhuang mural was the first one copied by Chang Shuhang?", (b). ``
  Will the Yuan Dynasty inherit the tradition of the Song Dynasty?'').}
  \label{teaser}
\end{figure}

However, providing free access to digital artifacts does not ensure that 
virtual museums can sufficiently engage public audiences 
in appreciating cultural values. It is particularly challenging to get
them to read and understand a large amount of information 
in an uncontrolled online environment \cite{Schweibenz1998TheM}. 
Although existing features such as 3D navigation \cite{KIOURT2016984},
content searching \cite{skov2014museum},
virtual reality (VR) \cite{Cheng:2017:TLC:3025453.3025857}, etc., 
enhance their sense of immersion, audiences in a virtual museum may still feel disoriented among the 
myriad of contents, not knowing what would be interesting for them \cite{bonis2009platform,SUN20131242,Rayward1999}. 
In the context of physical museums, researchers and practitioners 
show that a good tutor can encourage visitors to interact more with exhibits 
and understand them better
by asking guiding questions at a proper time, in a proper way \cite{HarvardMuseumsofNaturalHistory}. Even 
displaying inspiring questions without answers can spark
visitors' curiosity, and attract them to stay longer \cite{Roberts:2018:DEL:3173574.3174197}.
Figure \ref{teaser} (a) presents a real-world example, where museum staff posts questions
at the entry of a recent exhibition on Dunhuang Caves to engage visitors \cite{digitaldunhuang}. 


Inspired by these successful practices in physical museums, 
we propose to engage audiences in virtual museums through interactively prompting guiding questions
(Figure \ref{teaser} (b)).
In this paper, we use virtual museums of Chinese cultural artifacts as a case to examine 
our proposed approach. We aim to answer the following research questions (RQs):

\begin{enumerate}
  \item RQ1. How can we construct guiding questions that are more likely to engage audiences in virtual museums?
  \item RQ2. How would prompting guiding questions influence audiences' behaviors?
\end{enumerate}

To address RQ1, we first conducted interviews with experts to 
collect their opinions on virtual museums and how to engage audiences.
Then we analyzed cultural artifact-related posts shared in 
a popular question and answer (QA) community to
derive how to construct questions that are likely to 
attract general audiences in a museum context \cite{anderson2001taxonomy} .
Meanwhile, in terms of the proactive level for attracting users' attention \cite{BOWMAN2010927},
we designed two interaction mechanisms of prompting guiding questions: active and passive.
The active mechanism prompts questions when audiences watch the exhibit, which may distract 
their attention. The passive one prompts questions after 
audiences watch the current exhibit to reduce interruptions. 
To investigate the efficacy of our question and mechanism design on audience engagement (RQ2),
we conducted an online between-subject experiment with 150 participants.
Results showed that having interactive guiding questions encourages users to browse significantly more 
exhibits and considerably improve content comprehension. Interestingly, passive prompting got 
visitors to go both farther and deeper in virtual museums, 
while active prompting only leaded them to go farther. Finally, we invited
another 16 participants to compare the two mechanisms and share their thoughts and preferences. 
It seems that application questions work particularly well in the active setting, 
because they do not require higher level thinking that may distract audiences'
attention \cite{BOWMAN2010927}. In contrast, analysis, evaluation, and creation 
questions are more effective in the passive setting, due to the cognitive process for sparking
curiosity \cite{anderson2001taxonomy,litman2005curiosity,loewenstein1994psychology}. 
The contributions of this paper are:

\begin{enumerate}
  \item Through interviews with experts and the analysis of a social QA community, we extracted guidelines for 
  how to construct questions to attract online audiences.
  \item We designed two interaction mechanisms for prompting guiding questions.
  Through an online experiment, we showed the efficacy of using interactive guiding questions to 
  engage audiences in virtual museums.
  \item Through a qualitative comparison study,
  we obtained insights about the influence of question category and interaction mechanism.
\end{enumerate}

The remaining paper is organized as follows. We first review the background and related works. 
Then we present the overall flow and connections of our studies. 
Following that, we present each study and their findings in more details,
including domain expert interview, 
QA community analysis, interaction mechanism design, online experiment of guiding questions, 
qualitative comparison of the two interaction mechanisms. Finally, 
we discuss the design considerations derived from this study and limitations of this work.

\section{Background and Related Works}



\subsection{Museum and Virtual Museum}

A museum is \emph{``a non-profit, permanent institution in the service of society and its development,
open to the public,
which acquires, conserves, researches, communicates and exhibits the tangible and intangible heritage
of humanity and its environment for the purposes of education, study and enjoyment''} \cite{museumdefinition}.
The definition of museums given by International Council of Museums (ICOM) \cite{museumdefinition} 
shows the main responsibility of museums: protecting and promoting the sense of cultural heritage to
the public \cite{museumdefinition,SUN20131242}.
The extension of physical museums to virtual museums continues the mission
\cite{SUN20131242,10.1007/978-3-319-58559-8_12}.
Moreover, putting digitalized cultural artifacts online have multiple benefits.
From museums' perspective, it alleviates the congestion problem
in exhibition halls, protects and communicates cultural artifacts better \cite{TheCountenanceofPublicCulturalFacilities}.
For instance, the Palace Museum has an average visitors about 80,000 per day,
which results in great challenges for managing museums and provides
personalized visiting and learning experience \cite{TheCountenanceofPublicCulturalFacilities}.
From visitors' perspective, people can access digital artifacts without the limitation of
geographical locations and time. In addition, with the development of new
media technology, such as low-cost VR devices,
new experiences can be created in an affordable way \cite{novati2005affordable}.
The recent development of virtual museums aims to build museums in the cyberspace \cite{Schweibenz1998TheM},
where digital artifacts are not limited by a specific region or culture.
For example, the project Google arts and culture \cite{6691673}
integrates digital artifacts from different museums,
and forms a large scale online platform for people to explore.



\subsection{Practice in China}

In China, with the development of economy, people pay more attention to
cultural related activities recently. As explained in \cite{romero2014scholar},
this is because China is a diverse country in terms of culture, ethnic group,
and language, people are interested in learning and communicating
different cultural experiences. The efforts from the government is also huge. 
For instance, as reported in \cite{romero2014scholar},
more than 50 museums are built in the past 10 years in a small county.

In terms of virtual museums, the development in China can be divided into three stages
\cite{wenchangli}: online museums, digitalization of physical museums, and virtual museums.
Starting from the 1990s, some museums
began to build their websites spontaneously to publish activity news,
which is the first stage: online museums \cite{wenchangli}.
In the second stage, museums gradually digitalized their artifacts in a small scale, and putted
the digital content online for people to browse. Finally, the virtual museum
stage is still on-going, which aims to build museums in the cyberspace.
For virtual museums, the digital content will not be limited by one museum, but the whole knowledge space 
all over the world. For example, the digital project of Dunhuang academy \cite{dunhuangacademy}
aims to collect and organize all information about Dunhuang, not limited by the physical space.
Leading by the Palace Museum, many museums start
digitalizing their tangible artifacts, and put digital content online,
such as online exhibitions, video lectures, documentaries, etc.
\cite{TheCountenanceofPublicCulturalFacilities}. 
Despite the phenomenon, displaying plain collections can hardly draw people's interests 
and guide them properly \cite{10.1007/978-3-319-58559-8_12,6691673}. Therefore,
engaging audiences to understand and appreciate cultural values 
in virtual museums is still a challenging problem.

\subsection{Engaging in Museums and Virtual Museums}

In general, to engage visitors in museums or virtual museums, two approaches
can be investigated: increasing the sense of immersion and providing guidance.
For example, in physical museums, to increase the sense of immersion,
various new media technologies have been investigated in the past \cite{4160276,Muntean:2017:DCV:3025453.3025908}.
In \cite{4160276}, a mobile augmented reality (AR) application is proposed
to help people locate artifacts in physical museums. Visitors can explore 
museum space by interactively connecting information in mobile devices and 
artifacts in real world. In \cite{Muntean:2017:DCV:3025453.3025908}, Reese et al. investigate 
a storytelling mechanism to engage visitors, where users learn cultural values by exploring  
real artifacts in the physical space and media content in a tablet platform.
Similarly, in virtual museums, various technical approaches to increase immersion have been applied 
\cite{KIOURT2016984,Cheng:2017:TLC:3025453.3025857,skov2014museum}. 
But because of the short of exploring opportunities in a physical space,
engaging audiences in virtual museums is more challenging \cite{10.1007/978-3-319-58559-8_12,6691673}.

Although increasing the sense of immersion can engage visitors, to help them understand
and appreciate cultural values, proper guidance is needed 
\cite{mctavish2006visiting,10.1007/978-3-642-15892-6_30,Rayward1999}.
Empirical experience from
experts suggests that personalized tutors that conduct
conversations through asking questions can help engage visitors to appreciate the cultural value better
\cite{HarvardMuseumsofNaturalHistory}.
For instance, in \cite{10.1007/978-3-642-15892-6_30}, Swartout et. al show that 
using virtual tutors with language interaction can 
help children learn science and technology in a physical museum. 
And more recently, in a physical museum,
referring to psychology theory of curiosity \cite{litman2005curiosity,loewenstein1994psychology},
Roberts et. al. show that displaying big questions
can draw visitors' attention and increase their engagement in cultural artifacts effectively
\cite{Roberts:2018:DEL:3173574.3174197}. Successful practices of using guiding questions to engage 
visitors in physical settings imply the possibility to engage audiences in virtual museums.
But in terms of how to construct questions and how to interact with users in an online uncontrolled
environment, more care should be taken.  

\subsection{Question Study}

Asking questions is usually an important behavior for people to seek information.
Moreover, a selective question asked at a proper time, in a proper way can also mean more.
Existing research works show that
using questions can potentially attract humans' attention
\cite{Roberts:2018:DEL:3173574.3174197},
inspire creative thinking \cite{NIPS2017_6705},
provoke informative answers \cite{Hawkins2015WhyDY},
and lead to in-depth comprehension \cite{cohen1976learning}.
For constructing questions computationally,
one approach is data-driven
\cite{P17-1123}. But it suffers asking factual knowledge questions instead of creative ones \cite{P17-1123}.
Another approach formulates it as an optimization problem,
and constructs the best question based on the given context \cite{NIPS2017_6705}.
The Rational Speech Act (RSA) framework models question asking and answering
as a Bayesian reasoning process \cite{frank2012predicting,Hawkins2015WhyDY},
and can provide reasonable explanations of question construction. 
But similar to the optimization one, it is domain specific 
and not appropriate for inspiring creative and in-depth thinking in the context of cultural artifacts
\cite{NIPS2017_6705,frank2012predicting,Hawkins2015WhyDY}.

As the interest of Human-Computer Interaction (HCI), we study the perception of
questions from users' perspective. And in terms of cultural content appreciation,
we investigate how to ask a question
that can engage listeners in appreciating the
cultural value. In particular, we analyze existing
social QA platforms to find the patterns of what is a good question to engage audiences.

\subsubsection{QA Community}

The emerging social QA communities, such as Quora \cite{quora},
Zhihu \cite{zhihu}, etc.,
provide platforms for people
to share knowledge and experience.
In QA websites, audiences can ask questions, answer others' questions,
or watch others' questions and answers.
If a question is more interesting, people are more likely to view and answer it.
Analysis of QA communities can potentially provide insights on how to engage ordinary
people in appreciating cultural content. 
%

\section{Overall Flow of the Study}


\begin{figure}[h]
  \centering
  \includegraphics[width=0.9\linewidth]{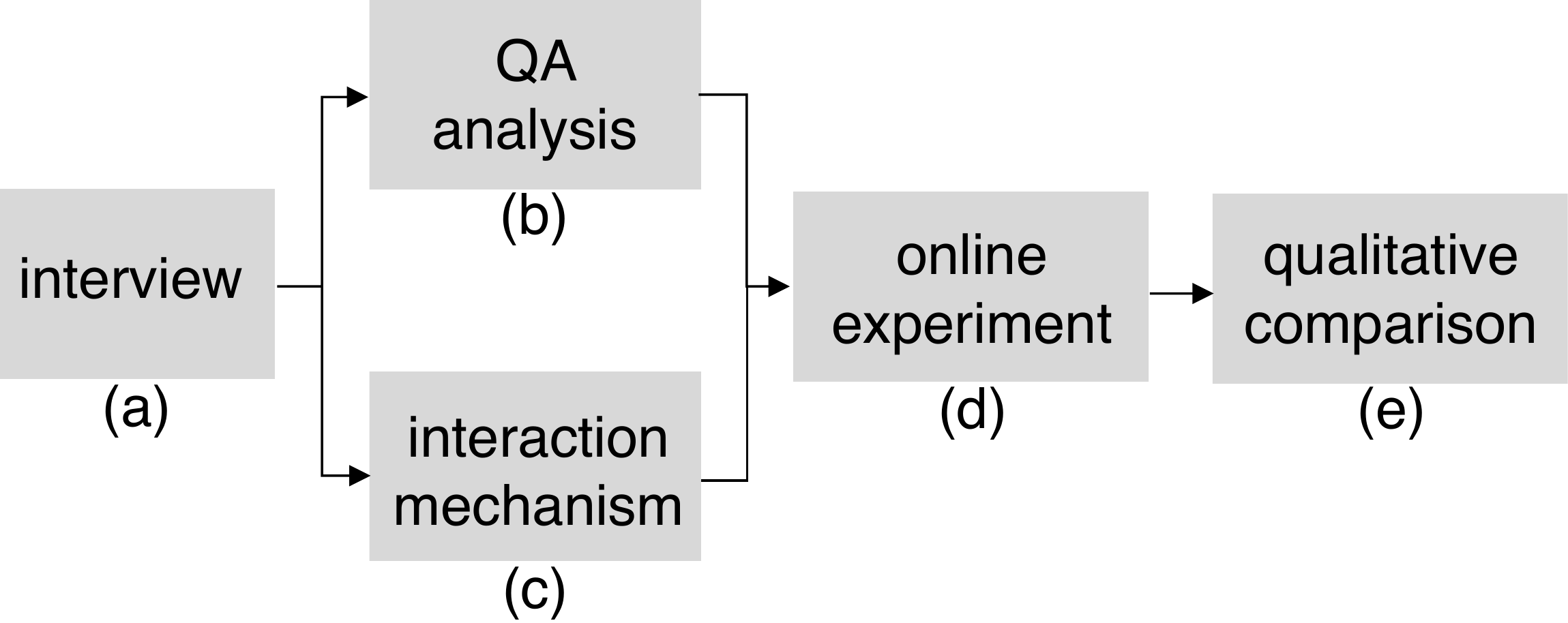}
  \setlength{\abovecaptionskip}{-2pt}
  \caption{The overall flow of the study in this paper.}
  \label{flow}
\end{figure}

\vspace{-10pt}

The overall flow of the study is summarized in Figure \ref{flow}.
In particular, we started from domain expert interviews (a) to obtain opinions of engaging 
audiences in virtual museums. 
Then we sought a popular social QA community Zhihu \cite{zhihu}, 
which is the largest Chinese QA community, 
to explore how to construct questions to engage visitors (b).
Meanwhile, in terms of the proactive level for attracting users’ attention \cite{BOWMAN2010927}, 
we designed two mechanisms to interactively prompt questions: active and passive (c).
To see the effects of guiding questions, we conducted an online experiment (d).
In particular, we built two online exhibitions, 
and equipped them with or without guiding questions.
Then we ran a between-subject experiment, and analyzed the results quantitatively.
Furthermore, we conducted another qualitative comparison to obtain more insights in terms of
question category and interaction mechanism (e). For all studies, 
except interviewing one expert in English,
we conducted in Chinese, and translated scripts into English for discussion and summarization.

\section{Domain Expert Interview}\label{interview}

To seek suggestions of engaging audiences in virtual museums from professionals,
we conducted semi-interviews with two domain experts of organizing museum exhibitions (E1, E2).
Each interview session lasted for about one hour. 

E1 is an executive officer in our university library,
and is in charge of organizing exhibitions in the university.
He presented his experience of organizing a \emph{Calligraphic Art}
exhibition in the university. The interview was conducted in English and on campus.
E2 is a team leader of cultural relics protection service 
in Chongqing, China, who has experience of conducting several online and offline exhibitions for children.
During the interview, E2 introduced her suggestions on engaging audiences in online exhibitions.
The interview was conducted remotely through instant messages.
We summarize our main findings in the following.


\subsection{Make Content Interesting and Easy to Understand}

E2 suggested to design exhibitions 
from children's point of view to make the exhibition easy to understand and interesting.
General audiences have different backgrounds, educational levels, 
and may not have enough expertise of the exhibition to 
understand various concepts, terminologies, and so on. So it is important 
to make the content of online exhibitions ``understandable''. 


\subsection{Help Audiences Memorize Some Keywords}

Both E1 and E2 confirmed the importance of making audiences learn from exhibitions.
E2 suggested to extract key elements of the exhibition.
And based on the elements, it is possible to derive questions to help audiences memorize
keywords, which encourages them to ``recall'' the exhibition.  



\subsection{Leave Time for Audiences}

Overwhelmed in a large amount of information, audiences may feel disoriented easily. 
As E1 introduced his experience of the combination of online and offline settings
to leave time for audiences to digest the content.
\emph{``So if they find some interesting things in the gallery,
they can further find more details on our ``exhibits'' web page."}

\subsection{Interact with Audiences}

E1 and E2 thought that interacting with audiences is the key to engage them.
In this way, people feel they are communicating with the exhibition. 
For example, E1 introduced his experience of inviting artists to provide ``face-to-face talks,
demonstrations, guided tours, etc.'', to deepen audiences' learning about exhibitions.
``In this way our exhibitions engage and interact with visitors.''

Learning from the interview, we can see that to engage audiences, both 
content and interaction are important. The content should be ``understandable''
and ``interesting'', and we should ``interact'' with audiences and ``leave time'' for them to think.
And as pointed out by E2, prompting questions can help audiences memorize ``keywords'' of the exhibition. 
Therefore, we consider two dimensions for designing our guiding questions: 
what types of questions and how to interactively prompt questions.
To explore what types of questions can attract audiences, we analyzed a popular QA community.
And we consider the problem of how to prompt questions from the aspect of attracting 
audiences' attention. More details are shown in the following sections.



\section{QA Community Analysis}\label{QACommunity}

We analyzed a popular social QA community Zhihu \cite{zhihu} to explore how to 
construct questions to engage audiences online.
In Zhihu, users can post, answer, and follow questions.
An attractive question can get more answers and followers.
Each question in Zhihu is classified into one or more topics, either done by
posters or dedicated editors. Therefore, a question can be seen as being
annotated into different categories.
We examined the hierarchy of all relevant topics about museums. After
regular group meetings and discussions, we chose the topic
\emph{cultural artifact} (\zh{文物}), and mined all questions in it.
In Zhihu, the topic \emph{cultural artifact} covers concrete
subtopics including Chinese painting, porcelain, bronzes, furniture,
building, etc., and also abstract ones like cultural heritage, preservation of cultural relics,
history of cultural artifacts, etc., which is suitable for our analysis purpose.

\subsection{Data Statistics}
Until May 20th, 2018, we collected 1041 questions under the topic \emph{cultural artifact},
with 3291 answers in total.
The average answer number is $3.6$ ($SD=15.2$), and the average follower number is $8.3$ ($SD=71.2$). 
The answer number and follower number are linear correlated (Pearson coefficient: 0.70, p-value: 0).
And the portion of questions that follower number is bigger than answer number is about $71.8\%$. 

\subsection{Question Category}

What types of questions may attract audiences and lead them to think?
We analyzed the most followed questions to find the pattern. 
In particular, we chose questions with follower number bigger than $5$ (about 
$19.8\%$ of all questions), and
ran the thematic analysis method \cite{braun2006using} to code the questions.
We used the Bloom's revised taxonomy of cognitive process
\cite{anderson2001taxonomy} as basic codes. The cognitive process
contains six categories to evaluate students' mastery of knowledge
in verb form: \emph{remember}, \emph{understand}, \emph{apply},
\emph{analyze}, \emph{evaluate}, and \emph{create}.  
Classifying questions into different cognitive categories can be seen
as having increasing level for inspiring thinking.
To annotate the questions, two authors first familiarized themselves with the crawled questions 
and the Bloom's revised taxonomy.
Following question construction guidelines of the Bloom's revised taxonomy \cite{question_construct,lord2007moving},
for each coding, we gave an argument to support the decision.
For instance, the argument of \emph{apply} could be reasoning with personal experience.


\begin{figure}[h]
  \centering
  \includegraphics[width=0.9\linewidth]{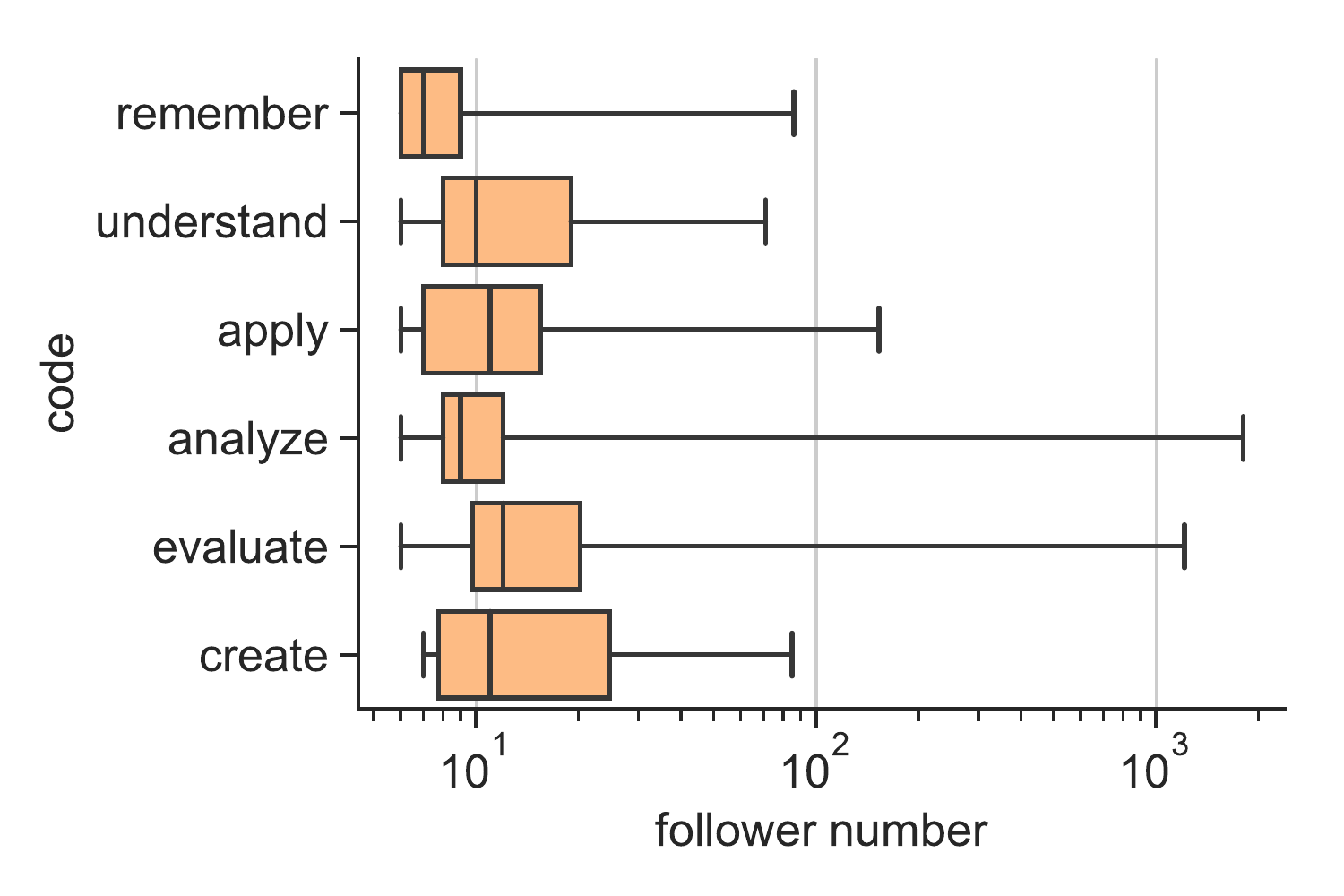}
  \caption{Box plot of the follower number of the coded questions (the follower number is in logarithmic coordinate).}
  \label{zhihu}
\end{figure}


The summarized box plot is shown in Figure \ref{zhihu}.
Generally, higher level questions in the Bloom's revised taxonomy can
draw more interests. For higher level questions that have only a few followers,
we found that they are usually lack of interesting key elements or short questions without too 
much background information, 
such as ``I want to study clinical medicine, but I am studying cultural relics protection. Is there any cultural relic protection direction associated with medicine?'' (\emph{analyze}, plain key element), 
``Is the Chinese clothing made by bamboo knots good?'' (\emph{analyze}, short question), and so on.
Therefore, for designing questions in our experiment,
we prioritized higher level questions with interesting 
key elements and background information.

\subsection{How to Construct Guiding Questions?}




Based on the thematic analysis, we summarized typical templates of
different question categories.
In addition, we also summarized common question features, such as what the question asks for, how many key 
elements, etc., as arguments for question construction. 
To make the templates easy to use, we first simplified the questions
by removing meaningless decorations and paraphrasing them to simple structures.
And we removed uncommon templates that have only a few examples.
Referring to the arguments and templates, new guiding questions 
in different categories can be constructed.



\subsubsection{Remember}

Questions in this category are usually short questions without any background information. 
For each question, there is only one key element, e.g., an unknown item, a place, cost, time, etc. 
For instance, ``what is Hanfu \footnote{the historical traditional dress of the Han Chinese}? ''
Several typical templates are:

\begin{itemize}
  \item What is ... ? 
  \item Where is ... ?
  \item How much is ... ?
  \item When is ... ?
\end{itemize}

\subsubsection{Understand}

Questions in this category 
ask for people' attitude, feeling, or thought about something. For example,  
``what kind of mood do most people like Hanfu have?'' Several typical templates are:

\begin{itemize}
  \item What is your/somebody's attitude towards ... ?
  \item What mood do/does ... have?
  \item What is your opinion about ... ?
\end{itemize}

\subsubsection{Apply}

People usually ask for an approach, personal experience, or an recommendation.
For instance, ``what kind of experience is having a puppet?'' Several templates are:

\begin{itemize}
  \item How ... ?
  \item Can you recommend ... ?
  \item What kind of experience is ... ? 
  \item What is ... used for?
\end{itemize}

\subsubsection{Analyze}

There are usually more than one key elements in questions.
To follow the questions, people need to do reasoning or inference.
For instance, ``why do we spend a lot of money to recover lost artifacts?''
Several templates are:

\begin{itemize}
  \item Why should we ... ?
  \item What is the reason ... ?
  \item Is ... the same to ... ?
  \item Is ... suitable/good/bad/true/false?
  \item What can we do if ... ?
\end{itemize}

\subsubsection{Evaluate}

Questions in this category ask for a judgment of a statement.
For instance, ``how do you think the school classifies Hanfu as fancy dress?''
Several templates are:

\begin{itemize}
  \item How do you think ... ?
  \item How to evaluate ... ?
  \item ..., is it true/false?
\end{itemize}

\subsubsection{Create}

In this category, people usually imagine some scenarios, and ask for potential results.
For example, ``imagine building a museum belonging to our generation hundreds of years later, what will be there?''
Several templates are:

\begin{itemize}
  \item Imagine ..., what will be ... ?
  \item If ..., can ... ?
  \item Will it be ... if ... ?
\end{itemize}



\section{Interaction Mechanism}\label{mechanism}

According to the domain expert interview, interaction is an important element to engage audiences. 
We consider it from the attention aspect \cite{BOWMAN2010927}, namely, when a question is prompted,
whether audiences concentrate all their attention to it, or part of attention.
In particular, we designed two interaction mechanisms: active and passive.
The active mechanism prompts questions more proactively when audiences watch the exhibit, 
and audiences need to multitask between watching the exhibit and a prompted question.
The passive one allows audiences to pay more attention to the exhibit, and only prompts 
questions when they change exhibits. 

\begin{figure}[h]
  \centering
  \includegraphics[width=1.0\linewidth]{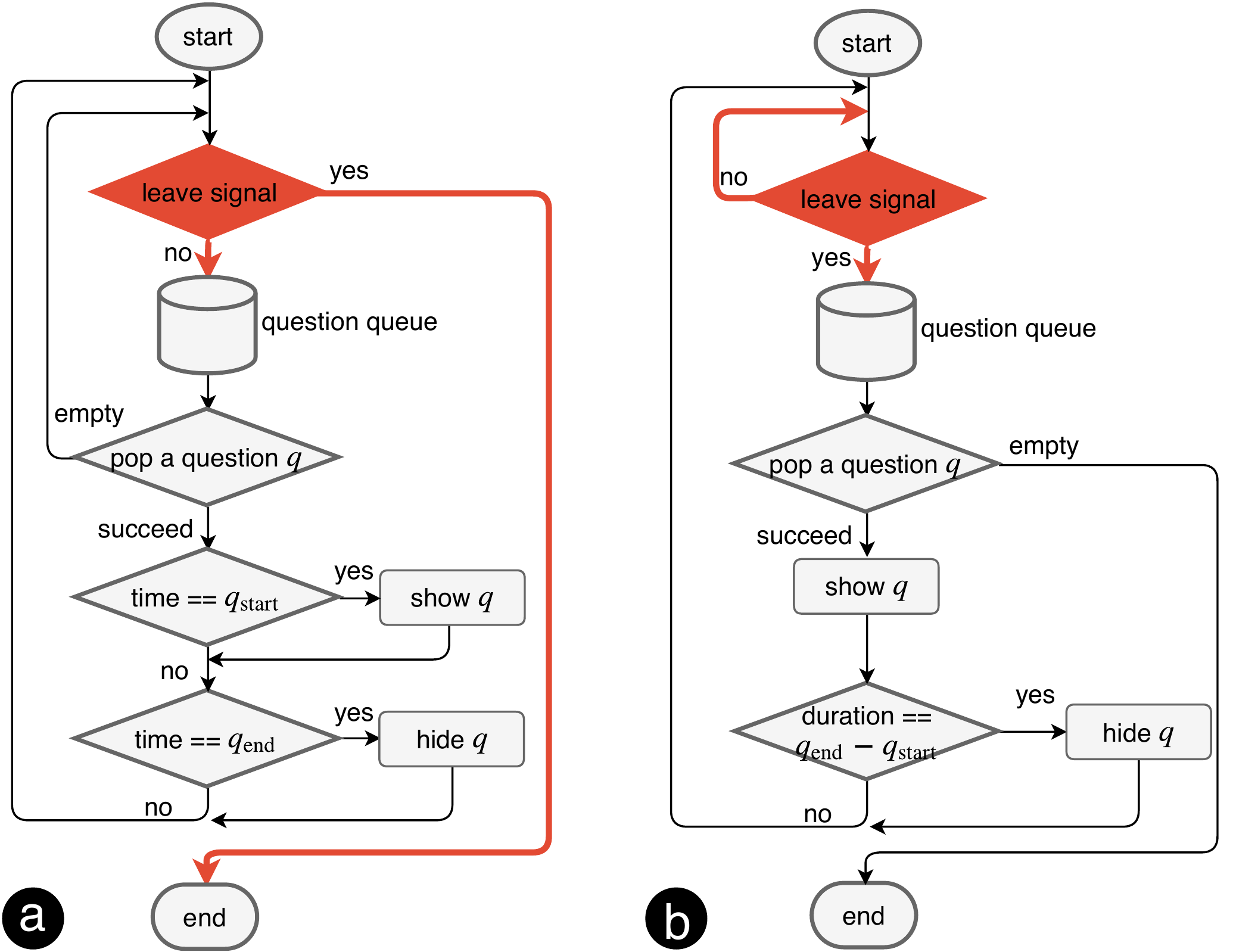}
  \caption{The flowcharts of the active mechanism (a) and the passive mechanism (b). Each question has 
  a prompting start time $q_{\text{start}}$, and an end time $q_{\text{end}}$.}
  \label{active_passive}
\end{figure}

The flowcharts of the two mechanisms are shown in Figure \ref{active_passive}.
The main difference is at the interaction moment when users try to leave the current slide. 
The active mechanism prompts questions when users browse the current exhibit based on preset time slots of questions, 
but does not interrupt them when they try to leave. Therefore, audiences need to multitask.
Instead, the passive one does not 
interrupt users during browsing the current exhibit,
but users need to watch all the questions before they change exhibits.
In this way, they can pay more attention to questions and exhibits.

For implementation, a question $q$ has a preset start time $q_{\text{start}}$ and a preset end time
$q_{\text{end}}$. For the active mechanism, a question will be prompted 
at $q_{\text{start}}$, and be hidden at $q_{\text{end}}$. For the passive mechanism,
the start prompting time of a question $q$ depends on users' input, and 
once $q$ is prompted, it will be hidden after the period of $q_{\text{end}}-q_{\text{start}}$.


\section{Online Experiment}\label{online_experiment}

To see how interactive guiding questions influence audiences' behaviors,
we conducted an online between-subject experiment with 150 participants. In particular,
we designed two online exhibitions, and equipped each with 
three versions of interaction mechanism: baseline, passive, and active. 
The passive and active version follow the design in section \ref{mechanism},
and the baseline version does not have guiding questions. 

\subsection{Online Exhibition Design}


We chose two unfamiliar topics: Chinese ritual bronzes \cite{chinesebronzes} and Chinese
painting \cite{chinesepainting}, and designed two online exhibitions in Chinese. 
The unfamiliarity can bring more challenges for engaging audiences.
For layout design, we referred to three typical online exhibitions \cite{fubaoshi,althani,feicui}
from the Palace Museum.
An online exhibition includes a menu bar for navigating into different themes, 
and under each theme, visitors can navigate a series of slides horizontally.


\begin{figure}[h]
  \centering
  \includegraphics[width=1.0\linewidth]{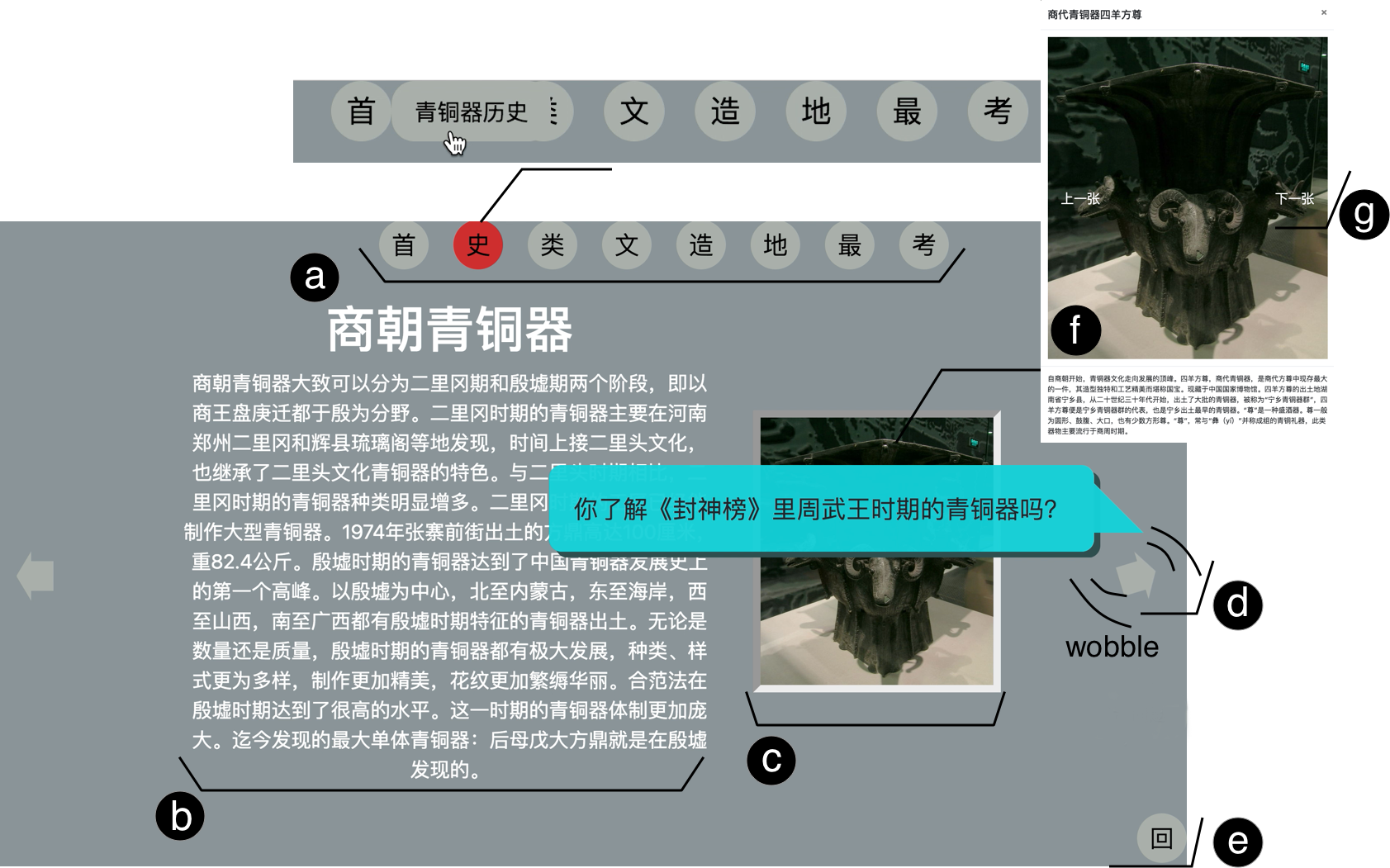}
  \caption{Typical layout of our online exhibitions: (a) menu bar, (b) text, (c) image asset,
  (d) navigation arrow, (e) exit, (f) dialog box, (g) navigation button.}
  \label{navigation}
\end{figure}


With similar complexity in terms of the number of captions and images, 
we designed the content of the two exhibitions.
The image assets and explaining text were downloaded from
Wikipedia \cite{chinesepainting,chinesebronzes} and searched results from Google that have proper copyrights.
The layout of the two exhibitions is the same.
As shown in Figure \ref{navigation}, users can
navigate the exhibition into different themes through the navigation bar (a). 
For each button, users can hover on it to see the full description.
Under each theme, there are several slides, and each slide contains text introduction (b)
and an image asset (c). When users click (c), a dialog box (f) will be popped up with 
more detailed introduction of the artifact.
Users can further view different artifacts through the navigation buttons (g).
The slides under a theme can be navigated through the left or right arrow (d).
When a guiding question is popped out, 
the position intended to be clicked by users will wobble for a moment to attract users' attention, such as 
the arrow (d) in Figure \ref{navigation}. Users can leave the exhibition by click (e). 
We refer to vertical complexity of an exhibition as the number of themes, 
and horizontal complexity as the maximal number of slides in different themes. In addition,
we refer to depth complexity as the maximal number of images contained in the image asset after clicking. 
The complexity of the Chinese ritual bronzes exhibition is $8$ for vertical, $7$ for horizontal,
and $5$ for depth. And the complexity of the Chinese painting exhibition is $5$ for vertical,
$20$ for horizontal, and $11$ for depth.

\begin{table*}
  \centering
  \begin{tabular}{ |C{4.2cm} | C{8cm} | C{4cm}|}
      \hline
      \small{Context} & \small{Question} & \small{Category}\\
      \hline
      \small{Painting of Luoshenfu (\zh{洛神赋})}
      & \small{Can you imagine how Cao Zhi's (\zh{曹植}) poem "Luoshenfu" could be expressed by painting?}
      & \small{Create} \\
      \hline
      \small{Inscription (\zh{铭文})} & \small{In general, bronzes with inscription are more precious. How you know think about inscription?}
      & \small{Understand} \\
      \hline
      \small{Ding (\zh{鼎})} & \small{How the ancients cook?}
      & \small{Apply} \\
      \hline
      \small{Erlitou culture (\zh{二里头文化})} & \small{What is Erlitou culture?}
      & \small{Remember} \\
      \hline
  \end{tabular}
  \setlength{\abovecaptionskip}{-5pt}
  \caption{Typical questions used in the two exhibitions.}
  \label{question}
\end{table*}

\vspace{-5pt}

\subsection{Question Design}

For each slide of the exhibition, we designed one or two questions that are used to guide users to browse
along vertical, horizontal, or depth direction.
Each question is at different position to inspire visitors to click,
including at the image position for depth direction, at the arrow position for horizontal direction, 
and at the menu position for vertical direction. 

Referring to the constructing guidelines in section \ref{QACommunity},
we wrote questions in different categories according to the slide content.
In particular, as suggested by E2 in section \ref{interview},
we first extracted key elements in the current slide, 
such as item, action, event, statement, and so on.
We selected one interesting element, 
and assigned its question category according to the arguments of different 
categories in section \ref{QACommunity}.
In addition, we prioritized higher level questions in the Bloom's revised taxonomy
and detailed background information to inspire thinking.
After assigning the category, we chose a suitable template in that category and constructed a question 
with the key element.
After multiple rounds of designing, discussing, and revising,
we created 135 questions in total for the two exhibitions (\emph{remember}: $21.5\%$,
\emph{understand}: $7.4\%$, \emph{apply}: $39.3\%$, 
\emph{analyze}: $19.3\%$, \emph{evaluate}: $2.2\%$, \emph{create}: $10.4\%$).
For comparing the effects in the following studies,
we further categorized them into three levels for inspiring thinking: low (\emph{remember},
\emph{understand}), middle (\emph{apply}), high (\emph{analyze}, \emph{evaluate},
\emph{create}), similar to the classification in \cite{lord2007moving}.
Low level questions only ask people to recall, interpret something, which needs 
the lowest level thinking. Middle level questions need people to apply one concept to another,
which needs middle level thinking. High level questions require more complex reasoning,
evaluating, creating, etc., which needs the highest level thinking.  
Several typical questions are shown in Table \ref{question}.

\vspace{-10pt}

\subsection{Experiment}

We conducted a between-subject experiment to minimize the learning effects \cite{lazar2017research},
and measured participants' browsing behavior and responses with the designed online exhibitions.  
During the experiment, each participant interacted with one version of two online exhibitions.
And we counterbalanced the order of the two exhibitions. We treated different guidance versions
as independent variable, and evaluate them in terms of 
the browsing behavior, content comprehension, and exhibition experience.  
We hypothesize that:


\begin{enumerate}
  \item \textbf{H1}. The guiding questions (passive and active) will encourage audiences browse virtual museums more.
  \item \textbf{H2}. The guiding questions (passive and active) will help audiences comprehend the content better.
\end{enumerate}


Whether having interactive guiding questions improves user experience is hard to predict.
In addition, without users' feedback, 
the exhibition experience of the two interaction mechanisms is also hard to compare.
We explore them through the experiment process, and further compare the two mechanisms 
through a qualitative study. 


\subsubsection{Participants}
We recruited participants by sending advertisements through platforms including a Chinese
survey platform WJX \footnote{\url{www.wjx.cn}} 
and crowdsourcing communities in QQ 
\footnote{QQ is a popular Chinese instant messaging application,
and people can create interest groups in it. The website is \url{im.qq.com}.}. Each participant can get 
2 CNY for about $0.5$ hours work.
Alternatively, we help them fill out their questionnaires as an exchange for their 
participation to our experiment. 
%
%
In total, 150 participants took the study (86 females, average age 21.8, SD: 5.65).
Each version contains 50 participants.


\subsubsection{Procedure}
We first showed an introduction webpage to participants about the procedure of the overall experiment,
following a questionnaire with six questions to test their 
background knowledge about the online exhibition content.
Each participant then browsed two online exhibitions. We counterbalanced
the order of the two exhibitions and recorded the browsing behavior.
After each exhibition, participants needed to fill out a questionnaire to report their
browsing experience. 
At the end of the experiment, participants were asked to fill out another questionnaire
with 16 questions to test their comprehension of the content.
For some participants recruited through QQ, we also conducted an interview
to ask their feedback about the exhibition and question experience (32 participants).












\subsection{Analysis and Results}

One-way MANOVA shows that there is a statistically significant difference 
in terms of of browsing behavior, content comprehension, and exhibition experience, 
$F(32, 264) = 1.66$, $p<0.05$; Wilk's $\Lambda=0.693$, $\eta^2=0.168$.  
We summarize the detailed statistical analyses 
in the following.

\subsubsection{Browsing Behavior: Click}

We count the click number of the two exhibitions as
$\sum_{e}\sum_{i_e,j_e,k_e}\delta_{i_e,j_e,k_e}^e$,
where $\delta=1$ if users view this page, otherwise $\delta=0$, 
$e=\{1,2\}$ is the index of the two exhibitions,
$i_e,j_e,k_e$ denote the indices of horizontal, vertical, and depth of the $e$ exhibition. 
The breath click number can then be denoted as $\sum_{e}\sum_{i_e,j_e}\delta_{i_e,j_e}^e$,
and the depth click number is $\sum_{e}\sum_{k_e}\delta_{k_e}^e$. To test whether the guiding questions
influence users' browsing behavior, we define
$\sum_{e}\sum_{i_e,j_e,k_e}\mathbb{I}_{i_e,j_e,k_e}^e$ as the guidance count,
where $\mathbb{I}$ is the indicator, if the next page users clicked is the page that the question 
indicated, $\mathbb{I}=1$, otherwise, $\mathbb{I}=0$. Therefore, if the guidance count of 
the passive or active version is higher than the one of the baseline, it shows our guiding questions influence
users' browsing behavior.


\begin{figure}[h]
  \centering
  \includegraphics[width=1.0\linewidth]{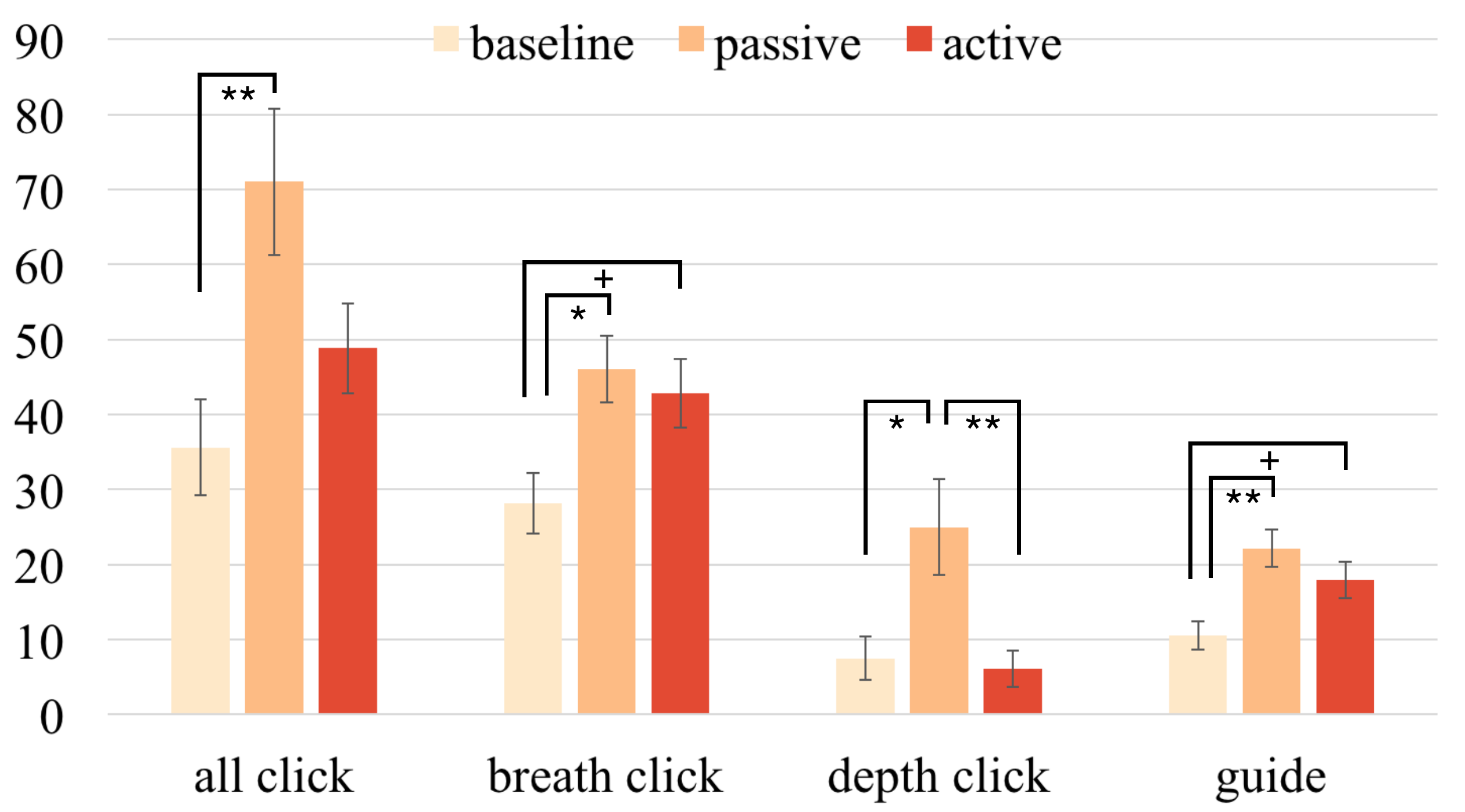}
  \caption{Means and standard errors of click number in terms of all pages, 
  breath direction, depth direction, and guidance count ($+: .05 < p < .1,*: p<.05,**: p<.01$).}
  \label{click}
\end{figure}


The summarized result is shown in Figure \ref{click}. One-way ANOVA analysis
shows that there is a significant effect of click number
($F(2, 147)=5.57$, $p<0.01$, $\eta^2=0.070$). In addition, 
Bonferroni post-hoc test shows that the click number
of the passive version is significantly bigger than the baseline version ($p<0.01$), but the 
the passive version and active version ($p=0.121$), and the active version and 
baseline version ($p=0.658$) are not significantly different.

There is a significant effect of breath click number
(one-way ANOVA, $F(2, 147)=4.80$, $p<0.05$, $\eta^2=0.061$). 
Bonferroni post-hoc test shows that the number of the
passive version is significantly bigger than the baseline version ($p<0.05$), and 
the active version is marginally bigger than the baseline version ($p<0.1$)
but the active version and passive version ($p=1.0$) are not significantly different.

Similarly, there is a significant effect of depth click number
(one-way ANOVA, $F(2, 147)=5.99$, $p<0.01$, $\eta^2=0.075$). 
Interestingly, Bonferroni post-hoc test shows that
the depth click number of the passive version is
significantly bigger than both the baseline version ($p<0.05$) 
and the the active version ($p<0.01$).
But the active version and the baseline version ($p=1.0$) are not significantly different.

The effect of guiding questions is significant 
(one-way ANOVA, $F(2, 147)=6.81$, $p<0.01$, $\eta^2=0.085$). 
Bonferroni post-hoc test shows that
the passive version is significantly bigger than the baseline version ($p<0.01$), and 
the active version is marginally bigger than the baseline version ($p<0.1$).
But the active version and the passive version ($p=0.561$) are not significantly different.


The result indicates our guiding questions generally can encourage users to click and browse more of the online 
exhibition content, specially for the passive interaction mechanism (\textbf{H1} partially supported).
In addition, it is interesting to notice that the active version seems not influence users' depth clicking 
behavior, while the passive version influence both depth and breath clicking. 
We will discuss it more in the qualitative comparison.

\begin{table}[h]
  \centering
  \begin{tabular}{|c|C{2cm}|C{2cm}|C{2cm}|}
    \hline
    & \small{low} & \small{middle} & \small{high} \\
    \hline
    \small{active} & \small{$6.7$} & \small{$\textbf{10.0}$} & \small{$9.3$} \\
    \small{passive} & \small{$10.1$} & \small{$10.8$} & \small{$\textbf{12.0}$} \\
    \hline
  \end{tabular}
  \setlength{\abovecaptionskip}{-5pt}
  \caption{The summarized ratios of the average click number per question category 
  (low: \emph{remember}, \emph{understand}; middle: \emph{apply};
   high: \emph{analyze}, \emph{evaluate}, \emph{create}).}
  \label{critical}
\end{table}

\vspace{-10pt}

\subsubsection{Browsing Behavior: Guidance}



We further investigate the browsing behavior in terms of question category and interaction mechanism together.
Because the number of different question categories is not equal and the positions are
not consistent, it is difficult to analyze it statistically. Therefore,
we examine it qualitatively by analyzing the guidance count in different question categories and 
interaction mechanisms.
In particular, we calculate the ratio of the average click number per question category
$\sum_{e}\sum_{i_e,j_e,k_e}\delta_{i_e,j_e,k_e}^e(c) / |Q_c|$,
where $\delta(c)=1$ if users view the page indicated by a question in the 
$c$ category, otherwise $\delta(c)=0$,
$|Q_c|$ is the number of question in the $c$ category.
We use the low, middle, and high categories defined previously,
and each category has roughly similar number of questions.
The result is shown in Table \ref{critical}. 

It is worth pointing out that different question categories have different effects on 
the interaction mechanism. Generally, higher level has a bigger ratio than the lower level,
which shows that questions that inspire people to think encourage them to browse more.
However, for the active version, the middle level seems to play a more 
important role than the high level. 
We give a detailed discussion in section \ref{qualitative_comparison}.


\subsubsection{Content Comprehension}

\begin{figure}[h]
  \centering
  \includegraphics[width=1.0\linewidth]{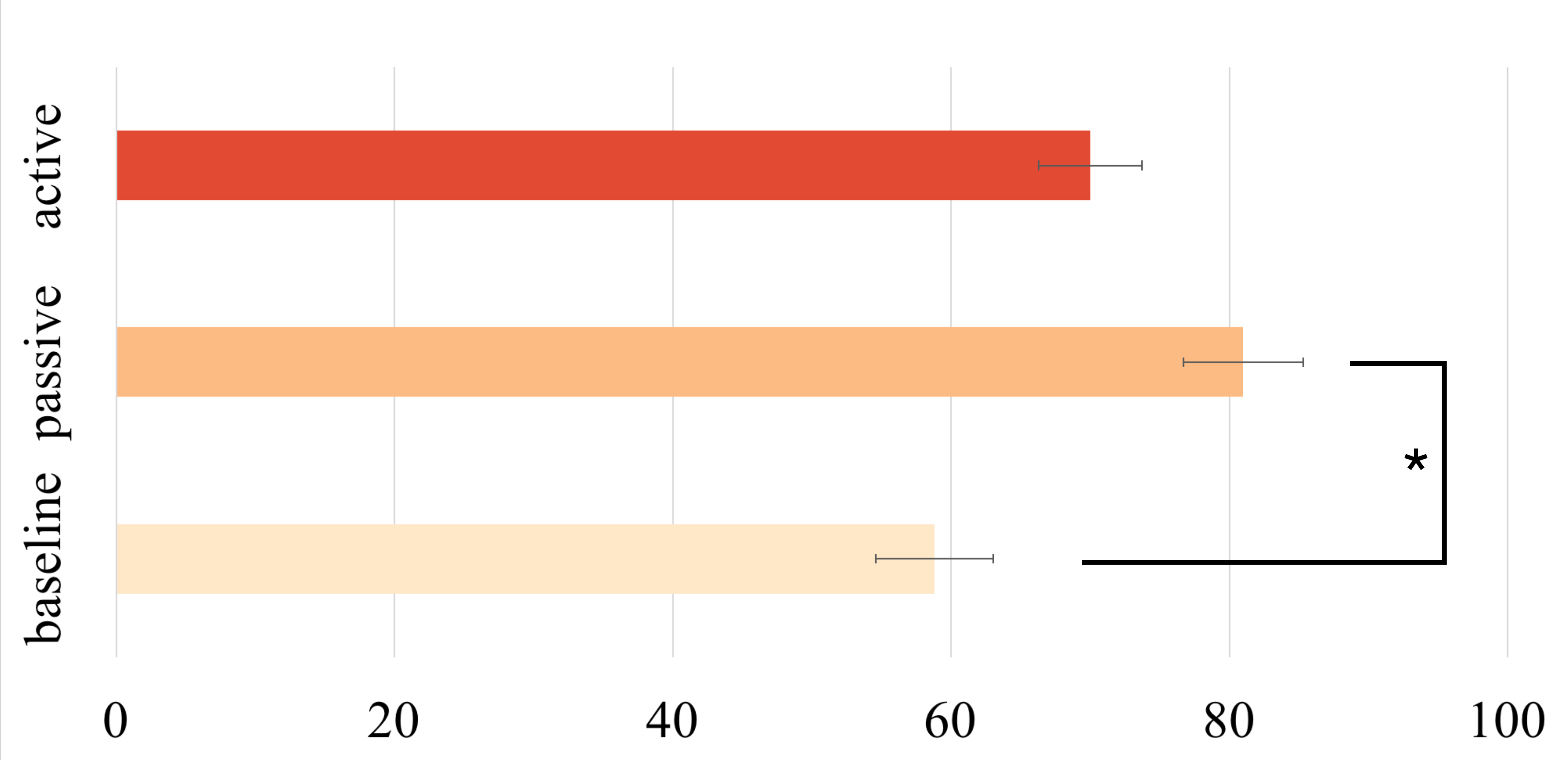}
  \setlength{\abovecaptionskip}{-10pt}
  \caption{Means and standard errors of post score ($*: p<.05$).}
  \label{score}
\end{figure}

One-way ANOVA analysis shows that no significant effect is found for 
the pre-testing score (baseline: $M=25.4$, $SD=11.47$,
passive: $M=28.6$, $SD=12.29$, active: $M=27.8$, $SD=12.17$,
$F(2, 147)=0.966$, $p=0.383$, $\eta^2=0.013$), which implies that
the samples of the three versions have similar background of the exhibition content. 
However, there is a significant effect of the post testing score 
($F(2, 147)=7.36$, $p<0.05$, $\eta^2=0.091$). 
In particular, as shown in Figure \ref{score},
Bonferroni post-hoc test shows that the score of the
passive version is significantly higher than the baseline version ($p<0.05$).
But the active version and the baseline version ($p=0.12$), the active version and 
the passive version ($p=0.241$) are not significantly different.
But there is a trend that the active version ($M=70.8, SD=26.41$) is 
higher than the baseline version ($M=58.8, SD=29.87$).
The result indicates that the passive mechanism can help general audiences comprehend and recall 
the exhibition content better (\textbf{H2} partially supported).
\emph{``I think the guiding questions are useful. Sometimes I just ignore some concepts.
But if it prompts a question to remind me, I would then think the answer and look at the text more 
carefully."--P120 (passive version)}.

\vspace{-10pt}

\vspace{-10pt}

\subsubsection{Exhibition Experience}

We measure the exhibition experience in terms of engagement, rewarding, 
satisfactory, and preference of the three versions on a 7-point Likert scale.
In particular, we ask participants four questions after each exhibition:
1). \emph{engagement}, please indicate the level you are engaged in the exhibition;
2). \emph{rewarding}, please indicate the level of rewarding after you browse the exhibition.
For example, after browsing the exhibition, 
you get to know some new knowledge, or you deepen the understanding of some concepts;
3). \emph{satisfactory}, please indicate the level you are satisfactory with the exhibition;
4). \emph{preference}, please indicate the level you like the exhibition.


One-way MANOVA analysis shows that there are no significant effects of the four measurements,
which indicates that the guiding questions do not interfere the 
exhibition experience significantly. However, there is 
a trend that the passive version makes people feel more engaged, rewarding, 
satisfactory, and preferable than the active version.


The non-significant effect may be due to users' different preferences.
From our interviews, the reported experience differs from individual to individual. 
For example, for the passive version, although most participants thought the questions 
can inspire them to read more of exhibits, 
others treated it as an interruption.
\emph{``It is not necessary to prompt questions for each slide, it is too annoying. Sometimes I just 
do not want to read (depth direction)."--P115 (passive version)}. We investigate it more in the following 
qualitative comparison section.





\section{Qualitative Comparison}\label{qualitative_comparison}

From the online experiment, we can see that the active and passive mechanism
influence users' behaviors differently. In addition, the interaction mechanism
seems to correlate with question category (Table \ref{critical}). 
To further compare the two designs and obtain insights on interaction mechanism
and question category,
we conducted a qualitative study with another 
16 participants (11 females, average age $21.1$, $SD=4.01$). 
The participants are recruited with crowdsourcing communities in QQ groups,
with a similar procedure in the previous experiment.
The average pre-testing score is $21.3$ ($SD=8.1$).
In this study, we asked each participant to first browse the two exhibitions
with the active and passive version separately.
The order of the two exhibitions and the two versions were counterbalanced.
We then asked participants to compare the two interaction mechanisms, 
and conducted a semi-structured interview with each participant
to collect their feedback.
The reported pros and cons of the two versions are summarized in Table \ref{preference}.

\begin{table}[h]
  \centering
  \begin{tabular}{|C{1.0cm}|C{2.85cm}|C{3.45cm}|}
      \hline
      & \small{pros} & \small{cons} \\
      \hline
      \small{active} 
      & \small{easy to interpret, 
      attention grabbing}
      & \small{interruptive} \\
      \hline
      \small{passive} & \small{leave time for browsing} 
      & \small{non-intuitive, require extra interaction (perceived)} \\
      \hline
  \end{tabular}
  \setlength{\abovecaptionskip}{-10pt}
  \caption{The summarized reasons for different preferences of the interaction way.}
  \label{preference}
\end{table}

\vspace{-5pt}

\subsection{Analysis of Interaction Mechanism}

\subsubsection{Interruption}

In many occasions, 
participants want to have a period of time to read the current content 
that they are interested in. Therefore, similar to our quantitative 
analyses in the previous study, most participants felt the passive version is better.
For example, some participants interpreted the guiding questions in
the active version as ``interruption''.
\emph{"I think prompting questions actively is too straightforward, and it interrupts me during 
reading. The second version (passive) is better. I can look at the content longer."--P3.}


\subsubsection{Input}

Multitasking between watching exhibits and guiding questions 
is more convenient for users if they want to see the questions. 
For example, some participants treated 
the questions as important clues to understand the exhibition. Therefore,
some participants interpreted the active version as ``intuitive'' and ``attention grabbing''.
\emph{``I like the first one (active version). The second one is too 
complex, not intuitive. The first one is better. It is more common, and shows me guidance more directly."--P5}.
In contrast, the passive version triggers guiding questions when users try to leave the current exhibit, 
which is perceived as needing ``extra operations''.

Prompting questions when users watch the current exhibit may 
interrupt them, which leads to poor comprehension (Figure \ref{score}). 
This aligns with previous work \cite{BOWMAN2010927} that multitasking lowers students' reading performance. 
However, due to the difficulty of determining when users finish the current exhibit, 
we may need extra operations to see guiding questions, as the passive version does.
This may possibly annoy users. Balancing guidance and interruption is critical for 
designing interaction mechanism of guiding questions.






\subsection{Analysis of Question Category}




\subsubsection{Align with Audiences' Understanding}

As mentioned in section \ref{online_experiment}, it is interesting that
middle level questions in the active version (\emph{apply}) encourage audiences 
to browse more than the high level ones (\emph{analyze}, \emph{evaluate}, \emph{create}, Table \ref{critical}).
From users' feedback, one possible reason is that 
questions in this category happen to align with audiences' understanding.
\emph{``I like the second one (active version), maybe because 
sometimes it prompts a question that happens to be the one I want to ask.
And when it appears in a proper time, it inspires me to read deeply."--P10}.
Because questions in this category need lower level thinking than high level questions,
the interruption brought by active interaction is not serious enough, i.e., people have time to digest the questions.



\subsubsection{Inspiring Thinking}

For the passive version, when people have more time to read the content, 
higher level questions play a more important role, as shown in Table \ref{critical}. 
In other words, if people have time to digest the content,  
it is more important to inspire them to think.
This may also explains the phenomenon in section \ref{online_experiment} 
that the passive version has a much bigger depth click number than others.  

The result confirms that the curiosity theory from psychology 
\cite{litman2005curiosity,loewenstein1994psychology} for engaging audiences
applies in online settings, which is consistent with previous works in physical museums 
\cite{Roberts:2018:DEL:3173574.3174197}. But designers should be more careful to 
deal with the relationship between curiosity and attention.

\section{Discussion}

In this section, we discuss several design considerations derived from our study 
and limitations of this work.

\subsection{Design Consideration}

\subsubsection{Use Interesting Language to Illustrate Content}

Making the content of virtual museums easy to understand is not enough,
it is better to use interesting language to attract audiences' interest.
For our online exhibition design, we used text descriptions from Wikipedia,
which are usually considered to be easy to understand. However, participants
in the online experiment still complained the content as ``dull'', ``not interesting''.
As suggested by E2, creating exhibits from children's perspective
could make it more appropriate to illustrate cultural content for public education.


\subsubsection{Guide Audiences When Necessary}

To change audiences' behaviors, interfering them when they watch exhibits
are unavoidable. Interestingly, our online experiment shows that the two 
interaction mechanisms do not influence participants' experience significantly.
Therefore, to encourage browsing and improve content comprehension,
it seems safe to use necessary interference methods to guide audiences.


\subsubsection{Maintain the Freshness of Interaction}

Although guiding questions do not harm the overall experience significantly,
interactively popping out questions in a uniform way for all slides still annoy users. 
For instance, in our qualitative comparison study, participants described ``always popping out 
questions'' as ``tedious'' and ``not necessary''.
To maintain the freshness of guiding questions, it is helpful to 
use different interaction mechanisms and switch between them.



\subsection{Limitation}

Our work has several limitations.
First, we only used Chinese cultural artifacts as an example, and conducted our experiment in China.
For generalizing the principles of question construction and interaction mechanism,
more systematic studies should be done.
Second, the starting and displaying time of a question in the active mechanism are fixed. 
If we can measure audiences' engagement level
\cite{Andujar:2013:LLE:2468356.2468480,Sun:2017:SHE:3025453.3025469}, and prompt 
questions accordingly, better experience can be expected.
Third, constructing guiding questions still needs manual work. 
Using the summarized guidelines to build a larger dataset and 
training question generation models \cite{du2017identifying,P171123,8424749}
can potentially deploy our method in a larger scale \cite{6691673}.
Fourth, our targeted audiences are ordinary people. 
For special groups like children, elderly, 
special interest visitors \cite{skov2014museum}, etc., more works should be considered.

\section{Conclusion}

We conducted a series of studies to understand how to interactively prompt guiding 
questions to engage audiences in virtual museums.
In particular, we used Chinese cultural artifacts as a case to 
examine our approach. We derived guidelines on how to construct questions to inspire different levels of thinking.
Through an online experiment and a qualitative comparison
study, we obtained insights about the influence of question category 
and interaction mechanism.
Further works could include automatic question construction, deploying and testing the 
method in a broader area.



